\documentclass[twocolumn,showpacs,preprintnumbers,amsmath,amssymb]{revtex4}

\usepackage{graphicx}
\usepackage{dcolumn}
\usepackage{bm}

\begin{document}

\preprint{APS/123-QED}

\title{A non-equilibrium dynamic mechanism for the allosteric effect}

\author{Jianhua Xing}
  \email{jxing@vt.edu}
\affiliation{%
Chemistry, Materials and Life Sciences Directorate, University of California and Lawrence Livermore 
National Laboratory, Livermore, CA 94550, and Department of Biological Sciences, 
Virginia Polytechnic Institute and State University, Blacksburg, VA, 24061-0406}

\date{\today}

\begin{abstract}
Allosteric regulation is often viewed as thermodynamic in nature. 
However protein internal motions during an enzymatic reaction cycle can be slow hopping 
processes over numerous potential barriers. We propose that regulating molecules may function by 
modifying the nonequilibrium protein dynamics. The theory predicts that an enzyme 
under the new mechanism has different temperature dependence, waiting time distribution of the turnover 
cycle, and dynamic fluctuation patterns with and without effector. Experimental tests of the theory are proposed.
\end{abstract}

\pacs{Valid PACS appear here}
\maketitle
A prominent property of enzymes (protein catalysts) is that their catalytic activities can be regulated. Enzymes that
are allosteric  
have two or more binding sites. Effects of ligand (effector) binding or reaction 
on one site can propagate to another distant catalytic site and affect its activity. Understanding the 
allosteric mechanism(s) is an important topic in structural biology. Conventional models assume that
effector binding modifies the equilibrium confomational distribution of allosteric proteins\cite{Monod1965}.
Recent proposed ``dynamic models" emphasize entropic (or the accessible configurational space)
 changes due to  effector-binding induced modification of protein fluctuation patterns  \cite{Volkman2001,
Formaneck2006,Fuentes2006,Swain2006,Hawkins2006}. Close examination 
reveals that these models actually 
share some basic ideas with the conventional models \cite{Wyman1990}. In summary  
allosteric regulation is generally believed to be ``fundamentally thermodynamic in nature"\cite{Wand2001}. 
For later discussions, we characterize 
these existing models as being driven by ``thermodynamic regulation". However, here we argue that the 
above thermodynamic description may be incomplete, and propose an alternative ``nonequilibrium dynamic regulation
 mechanism" . 
This idea is inspired by experimental and theoretical studies on dynamic disorder, the phenomena that the 
``rate constant" of a process is actually a statistical function of time due to slow protein conformational motions 
\cite{Austin1975,Zwanzig1990,Xie1999,English2006,Min2005}.\\
In this work we specifically examine allosteric regulation of enzymatic reactions. Furthermore, we consider
 the case of positive regulation (i.e., effector binding results in higher activity) unless specified otherwise.
 The catalytic site being regulated can be described by a few slow global conformational modes (here we assume one for
simplicity) and local conformational changes involving atomic rearrangement (here refering as the reaction coordinates). For barrier crossing processes, 
a system spends most of the time near the potential minima, and the actual barrier-crossing time 
is transient. Therefore, one can reduce the potentials further
 to one-dimensional projections along the conformational coordinate, 
and approximate transitions along the reaction coordinate by rate processes between the 
one-dimensional potential curves.
Similar description has been used in other contexts (e.g., protein motor studies \cite{Bustamante2001}). 
Protein dynamics is affected by substrate binding. A minimal
 model representing the states of a catalytic site is: E (empty), S (substrate bound), 
P(product bound). Fig. 1a illustrates an example used in this work. The protein states are 
described by potential curves along the conformational coordinate with localized transitions between them. 
For an enzymatic cycle, a substrate molecule first binds 
onto the catalytic site (E$\rightarrow$S), then forms a more compact complex from which a 
chemical reaction takes place (S$\rightarrow$P), and finally the product is released (P$\rightarrow$E). 
In the more familiar discrete kinetic form, the overall process can be represented as 
$E+S \rightleftharpoons ES \rightleftharpoons ES^* \rightleftharpoons EP \rightleftharpoons E+P$, 
with $E$, $S$ and $P$ refer to the enzyme, substrate, and product respectively. Notice that in 
general, the optimal conformational coordinates for reactant binding, the chemical reaction, 
and product release may not be the same (as also suggested experimentally \cite{Boehr2006}), and some 
conformational motion is necessary during the cycle. Dynamics of the reduced system can be described
 by a set of over-damped Langevin
 equations coupled to Markov transitions \cite{Zwanzig2001},
$\zeta_i \frac{dx}{dt}=-\frac{dU_i(x)}{dx}+f_i(t)$, 
where $\it{x}$ represents the conformational 
coordinate, $\it{U_i(x)}$ is the potential of mean force at a given substrate binding state, 
$\it{\zeta_i}$ is the drag coefficient, and $\it{f_i}$ is the random fluctuation force with the 
property $<f_i(t)f_i(t')> = 2 k_B T\zeta_i\delta(t-t')$, with $k_B$ the Boltzmann's constant, $T$ the temperature. 
Chemical transitions accompany motions along the conformational coordinate with 
$x$-dependent transition rates. For simplicity we leave the more general description, 
the generalized Langevin equation \cite{Min2005, Xing2006, Zwanzig2001}, for future studies.
The dynamics can be equally described by a set 
of coupled Fokker-Planck equations (here we only consider the steady state),
\begin{equation}
\label{FPequation}
-\frac{D_i}{k_B T}\cdot \frac{\partial}{\partial x} \left (
-\frac{\partial U_i}{\partial x}\rho_i\right )
+D_i\frac{\partial^2\rho_i}{\partial x^2}
+\sum_{j \neq i} (k_{ij}\rho_j-k_{ji}\rho_i) 
= 0
\end{equation}
Where $D_i = k_BT/\zeta_i$ the diffusion constant, $k_{ij}$ the transition matrix element, and $\rho_i$ the probability 
density to find the system at position $x$ and state $i$. For simplicity we dropped the dependence of $U_i$ and $\rho_i$ on $x$ in the above equation and later discussions. 
This is a unified framework for describing allosteric regulation. The existing models can be regarded as
special cases with the conformational coordinate discretized \cite{Wyman1990}.
\begin{figure}
\includegraphics{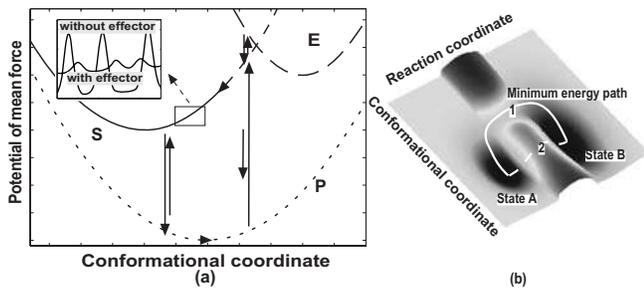}
\caption{\label{fig:epsPES} A minimal model for the catalytic site (a)The free energy curves represent 
three distinct catalytic site binding states (E, S, P) projected onto the conformational coordinate x. 
The inlet illustrates that the smooth potentials are actually coarse-grained over rugged potential surfaces.
 Effector binding at a distant site may modify the roughness of the potentials .  
(b) Two classes of reaction pathways along a multi-dimensional surface. Path 1 is along the
long but low barrier minimum energy path. Path 2 is a short high barrier corner-cutting path.}
\end{figure}
The ``thermodynamic regulation" models assume that effector binding at a remote site can affect the dynamics 
at the catalytic site by modifying $U_i$. While differing in details, these models assume a quasi-equilibrium 
distribution along the conformational coordinate, thus a thermodynamic treatment is appropriate \cite{Wyman1990}. 
However, protein conformational fluctuations can be very slow (e.g. from ms to minutes 
\cite{Volkman2001,Min2005,Boehr2006}), which is comparable or even slower than the enzyme turnover time.  
Consequently, protein fluctuations may not have fully accessed the conformational space, and thus not be 
in equilibrium. Variation of the dynamic properties along 
the conformational coordinate can have a dramatic effect on the apparent protein activity. 
Our recent theoretical analysis showed that the observed slow protein conformational dynamics 
can be explained by rugged protein potential surfaces\cite{Xing2006}. 
During relative motions between two protein parts, numerous noncovalent bonds(residue pairs with electrostatic, 
hydrophobic/hydrophilic, steric interactions, etc.)
 may form and break with associated local conformational changes.
These processes are in general uncorrelated with each other, which result in rugged potentials \cite{Stein1985}. 
The relative motions are then characterized by 
hopping over numerous potential barriers (refer to inlet of Fig. 1a). For a potential curve with
 random ruggedness, Zwanzig showed that the barrier-hopping process can be approximated by
 diffusion along a coarse-grained smooth potential with an effective diffusion constant 
$D=D_0\exp(-(\epsilon/k_BT)^2)$, where $D_0$ is
 the bare diffusion constant, and $\epsilon$ is the potential roughness parameter \cite{Zwanzig1988}. The reported value of
 $\epsilon$ is 2-6 $k_BT$ \cite{Nevo2005}. With $D_0 = 10^{-6} cm^2/s$, $D$ can be reduced to $1 \AA^2/s$ 
with $\epsilon \sim 4.8 k_BT$. Thus internal diffusion can be a rate limiting step for enzymatic 
reactions and in principle can be regulated by allosteric effects (see Fig. 1). 
Further studies are necessary to clarify the atomic view of the proposed potential roughness regulation.  
The effective diffusion can be accelerated by inducing 
local conformational changes and synchronizing the breaking and formation of the noncovalent bonds.
 It may be also related to the coupling mechanism between global and local vibrational
modes dicussed by Hawkins and McLeish \cite{Hawkins2006}. \\
%
In addition to being a theoretical possibility, the nonequilibrium regulation mechanism also has the following improvements
 over the conventional regulation mechanisms.\\
First, it is a more effective way to regulate the enzyme activity than 
the thermodynamic regulation (conformational and entropic) mechanisms. To increase the activity by $10^{10}$ through
 an Arrhenius process, the activation``free energy" barrier needs to be lowered by 23 $k_BT$. Similar amount of free energy change is needed
 for an equilibrium  population shift mechanism over the reactant conformational space. On the other hand, 
for an internal diffusion limited process, the reaction rate is linearly dependent on the effective 
diffusion constant. To increase the activity by the same $10^{10}$, the lower bound of the roughness
 parameter only needs to be adjusted by $\sim 5 k_BT$.\\
Secondly, compared to the conformational change mechanism, 
the nonequilibrium dynamic regulation mechanism has less requirements on the mechanical properties 
of the protein, similar to the proposed equilibrium dynamic models. 
The distance between the two binding sites of an allosteric protein can be far (e.g., 
15 nm for the bacterial chemotaxis receptor \cite{Kim2002}). Under the conformational change mechanism, effective coupling 
of the two sites requires a faithful finely tuned transmission of the mechanical strain due to ligand 
binding from one site to another one through a set of mechanical stress relaying network. 
These network residues must have mechanical properties distinctive from
 other residues to minimize energy dissipation to the surroundings. Otherwise, a significant portion
 of the effector binding energy would be wasted. In other words, coupling between these relaying 
residues and others should be minimized. By comparison, under the current nonequilibrium 
or the existing dynamics regulation mechanisms, 
the effect of effector binding can be highly nonlocal. Effector binding may 
affect the other site by finely regulating local structures far away from that site. By modifying the 
effective diffusion constant, these local modifications may affect the dynamics along the conformational
 coordinate $x$ in our formalism through coupling to global modes. 
The latters are composed of collective motions of residues within
 the catalytic site and those far from it. The effect manifests itself through larger root-mean-square 
deviation as observed in NMR, x-ray crystallography, and in molecular dynamics simulations 
\cite{Formaneck2006,Volkman2001,Kim2002}. 
The dynamics of the collective motions should be examined as well.\\
\begin{table}
\caption{\label{tab:table1}1 Model parameters. Here (E, S, P) refer to empty, substrate bound, and product bound, respectively. 
All energy units in this table are in reduced units, $k_BT= 1$. For simplicity,$D_E = D_S = D_P$. 
The prefactors of the rate constants are chosen so that the maximum values of $k_{SE}$, $k_{PS}$, and $k_{ES}$
 are approximately 10, 1.5, and 5, respectively.}
\begin{ruledtabular}
\begin{tabular}{ccccccc}
 &$S\leftarrow E$ &$P\leftarrow E$ & $E\leftarrow S$ 
 &$P\leftarrow S$ &$E\leftarrow P$ & $S \leftarrow P$\\
\hline
$k^0_{ij}$ & 2e2 & 2e-3 & 2e2 & 1.6e3 & 2e3 & 1.6e3 \\
$U_{ij}^{0\dagger}$ & 3 & 3 & 3& 6 & 3 & 6\\
$L_{ij}$ & 0.3 & 0.3 & 0.3 & 0.3 & 0.3 & 0.3\\
$x_{ij}^e$ & 0.65 & 0.65 & 0.65 & -0.65 & 0.65 & -0.65\\  
\end{tabular}
\end{ruledtabular}
\end{table}
In our numerical calculations, the potentials are chosen to be harmonic potentials (see Fig. 1a),
$U_i=\frac{1}{2}\kappa_i(x-x_{0i})^2+U^0_i$ , $\kappa_i = (1, 0.5, 0.4)$, 
$x_{0i} = (1, -0.5, 0)$, and $U^0_i = (0, -1, -3)$. To model transitions between 
different states, we also model the transition state potentials by harmonic potentials, 
$U_{ij}^{\dagger}=\frac{1}{2} ( ( x - x_{ij}^e ) / L_{ij})^2+U_{ij}^{0\dagger}$. The transition rate from state $j$ to $i$ is given by
$k_{ij}(x) = k_{ij}^0 \exp [(U_j(x)-U_{ij}^\dagger(x))/k_BT]$  with parameters given in Table 1.  
By solving Eq. \ref{FPequation}, the enzyme turnover rate (which measures how many substrate molecules can be
transformed into product by an enzyme molecule per unit time) is calculated by $r = \int (k_{EP}\rho_P-k_{PE}\rho_E)dx$.
Eq. \ref{FPequation} was solved with the algorithm developed by Wang $\it{et al.}$ 
at given values of $D$ and $T$ \cite{Wang2003}.  
The algorithm discretizes the conformational coordinate, and transforms the partial differential equations into a jump process over many discrete states 
with their normalized populations $p$ (defined as the probability density integrated over the discrete regions) described in the form
$\bf{Kp}=0$. The composite $\bf{K}$ matrix contains transitions along both the conformational and reaction coordinates(see the original paper for details).
Fig. 2a shows the calculated relative enzyme turnover rate as a function 
of the internal diffusion constant. While the diffusion constant is not a rate-limiting parameter at high $D$ values 
(as compared to the chemical transition rates), at smaller values of $D$ the turnover rate depends on the
 diffusion constant linearly which is a signature of the existence of diffusion-limited steps. Fig. 2b 
shows the temperature dependence of the turnover rate. With high values of $D$, the exponential $1/T$ 
dependence mainly comes from the Arrhenius dependence of the transition rates. However with small 
values of $D$, the turnover rate shows strong non-exponential dependence, since the effective diffusion 
constant $D$ has a Gaussian dependence on $1/T$. Experiments can test the predicted different temperature 
dependence of enzyme activity with and without the effector provided it is regulated by the current 
mechanism. Fig. 2c also shows the waiting time distribution between two consecutive turnover cycles 
calculated using the formula derived by Gopich and Szabo \cite{Gopich2006}, 
\begin{equation}
P(\tau)=\bf{1}^{\dagger} V 
             \exp (\int_0^\tau (K-V)dt) 
                V p/ (1^\dagger V p), 
\end{equation}
where $1^\dagger$ refers to vector contraction, $\bf{K}$ the above composite transition matrix,
 $\bf{V}$ the transition matrix with only product release transitions 
as the sole nonzero elements, $\bf{p}$ the obtained steady-state population distribution.
A system with low $D$ 
values shows non-exponential distribution due to dynamic disorder. At high $D$ values, effects of the dynamic
 disorder diminish and the distribution is exponential. Therefore, we predict that an enzyme functioning under
 the new dynamic regulation mechanism shows larger dynamic disorder effects. This can be  tested by 
measuring consecutive single enzyme turnover time distributions with and without the effector, an extension 
of the work done by the Xie group \cite{English2006}.\\
\begin{figure}
\includegraphics{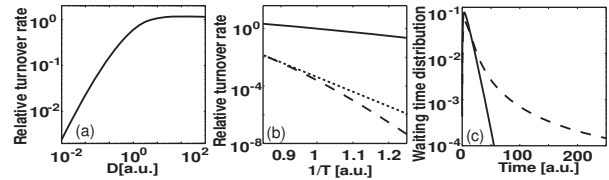}
\caption{\label{fig:epsResult} Theoretical predictions. (a) Enzyme turnover rate as a function of the effective internal 
diffusion constant $D$. (b) Temperature dependence of the enzyme turnover rate. $D=D_0\exp[-(\epsilon/k_BT)^2]$, 
where $\epsilon$ is the roughness parameter, and $D_0 = 10^3$. Solid line:$\epsilon = 0$. 
dashed line: $\epsilon = 4 k_BT$, the dotted line gives the exponential dependence 
if the Arrhenius rate equation is assumed. 
The temperature dependence of $D_0$ is neglected in this calculation. 
(c) Turnover waiting time distribution with $D = 1$ (solid line) and $D = 0.001$ (dashed line).}
\end{figure}
For an enzymatic reaction under allosteric regulation, effector binding changes its reaction rate from $k_1$ to $k_2$. 
Let's define an effective free energy barrier change 
$\Delta\Delta G^{\dagger} = k_BT\vert\ln(k_2/k_1)\vert$, which must be due to the effector binding energy. 
Is it necessary that the effector binding energy be no smaller than $\Delta\Delta G^{\dagger}$?  
 In a related question,
 for a system described by a multi-dimensional potential surface shown in Fig. 1b, is it possible that 
path 1 is dynamically comparable or even unfavorable than path 2? The answer to the latter is yes, provided that 
path 1 involves slow diffusion processes so that the average time for the transition along path 1 is even 
longer than path 2. A similar situation is discussed for tunneling pathways 
(e.g., proton-transfer reactions): the so-called corner-cutting large
 curvature tunneling and small curvature tunneling \cite{George1972}. In this work we discussed that slow diffusion within 
a protein is physically possible due to the rugged potential surfaces. In his barrier crossing theory, 
Kramers derived dependence of the barrier crossing rate on the barrier height, the drag coefficient, and other potential parameters 
of the system \cite{Kramers1940}. While most enzymatic reaction studies focus on barrier height changes, modification 
of the drag coefficient can affect enzyme activity (see also \cite{Hamelberg2005}). For proteins with rugged potentials, 
here we propose that the effective drag coefficient and thus protein activity can be tuned over a broad 
range by modifying potential roughness. In this case, the effector binding energy need not to be less than 
$\Delta\Delta G^{\dagger}$. \\
The dynamic mechanism discussed in this work is related to other studies discussing protein dynamic properties and 
allosteric regulation by considering the effect of rugged potential landscapes \cite{Swain2006}). However, 
there is also a fundamental difference. The current model treats the enzymatic reaction as a nonequilibrium problem in general. 
In their NMR studies of dihydrofolate reductase catalysis, Boehr et al. shows that the internal conformational 
motion is the rate-limiting step \cite{Boehr2006}.
This system may provide a nice test system for the proposed mechanism. In addition, slow conformational dynamics has been 
observed for allosteric proteins \cite{Volkman2001} further supporting the validity of our formalism. 
Some experimental observations supporting the existing dynamic models are also consistent with the current model. 
Kim et al. and Popovych et al.  proposed that the allosteric
 signal is transmitted through dynamical rather than conformational changes \cite{Kim2002,Popovych2006}.
We expect that the allosteric mechanism of a given protein has contribution from both thermodynamic
 regulation (the conventional conformational change mechanism \cite{Monod1965} 
and the newly proposed entropic effect  \cite{Volkman2001,
Formaneck2006,Fuentes2006,Swain2006,Hawkins2006}), 
and nonequilibrium dynamic regulation proposed in this work. Different proteins may differ on which effect is dominant. 
\begin{acknowledgments}
I thank Professors George Oster (UC Berkeley), Sung-Hou Kim (UC Berkeley), Hong Qian (U Washington), 
and Qiang Cui (U Wisconsin), Drs. Daniel Barsky, Ken Kim, Michael Surh, Todd Suchek at LLNL, 
Tongye Shen (UCSD), and Mr. Wei Min (Harvard) for helpful comments. JX is supported by a Lawrence Livermore National Laboratory Directed 
Research and Development grant, and by a Chemistry, Material, and Life Sciences Directorate fellowship. 
This work was performed under the auspices of the U.S. Department of Energy by the University of California, 
Lawrence Livermore National Laboratory under Contract No. W-7405-Eng-48.
\end{acknowledgments}

\begin{references}
\bibitem{Monod1965} J. Monod, J. Wyman, and J. P. Changeux, J. Mol. Biol. {\bf 12}, 88 (1965); D. E. Koshland, G. Nemethy, and D. Filmer, Biochemistry 5, 365 (1966).
\bibitem{Volkman2001} B. F. Volkman, D. Lipson, D. E. Wemmer, et al., Science 291, 2429 (2001); 
D. Kern and E. R. P. Zuiderweg, Curr. Opin. Struc. Biol. 13, 748 (2003).
\bibitem{Fuentes2006} E. J. Fuentes, S. A. Gilmore, R. V. Mauldin, et al., J. Mol. Biol. 364, 337 (2006); 
K. Gunasekaran, B. Y. Ma, and R. Nussinov, Proteins-Structure Function and Bioinformatics 57, 433 (2004); 
A. Cooper and D. T. F. Dryden, Eur. Biophys. J. Biophys. Lett. 11, 103 (1984); 
D. M. Ming and M. E. Wall, Proteins-Structure Function and Bioinformatics 59, 697 (2005); 
J. P. Ma and M. Karplus, Proc. Natl. Acad. Sci. U.S.A. 95, 8502 (1998); 
S. Jusuf, P. J. Loll, P. H. Axelsen, J. Am. Chem. Soc. 125, 3988 (2003);
R. J. Hawkins, T. C. B. McLeish, Phys. Rev. Lett., 93, 098104 (2004).
\bibitem{Swain2006} J. F. Swain and L. M. Gierasch, Current Opinion in Structural Biology 16, 102 (2006); V. J. Hilser, B. Garcia-Moreno, T. G. Oas, et al., Chem. Rev. 106, 1545 (2006).
\bibitem{Hawkins2006} R. J. Hawkins, T. C. B. McLeish, Biophys. J. 91, 2055 (2006).
\bibitem{Formaneck2006} M. S. Formaneck, L. Ma, and Q. Cui, Proteins-Structure Function and Bioinformatics 63, 846 (2006).
\bibitem{Wyman1990} J. Wyman and S. J. Gill, Binding and linkage: functional chemistry of biological macromolecules (University Science Book, Mill Valley, CA, 1990).
\bibitem{Wand2001} A. J. Wand, Science 293, 1395 (2001).
\bibitem{Austin1975} R. H. Austin, K. W. Beeson, L. Eisenstein, et al., Biochemistry 14, 5355 (1975).
\bibitem{Zwanzig1990} R. Zwanzig, Acc. Chem. Res. 23, 148 (1990).
\bibitem{Xie1999} X. S. Xie and H. P. Lu, J. Biol. Chem. 274, 15967 (1999).
\bibitem{English2006} B. P. English, W. Min, A. M. van Oijen, et al., Nat. Chem. Biol. 2, 87 (2006).
\bibitem{Min2005} W. Min, G. B. Luo, B. J. Cherayil, et al., Phys. Rev. Lett. 94, 198302 (2005).
\bibitem{Bustamante2001}C. Bustamante, D. Keller, G. Oster, Acc. Chem. Res. 34, 412 (2001).
\bibitem{Boehr2006} D. D. Boehr, D. McElheny, H. J. Dyson, et al., Science 313, 1638 (2006).
\bibitem{Zwanzig2001} R. Zwanzig, Nonequilibrium Statistical Mechanics (Oxford University Press, Oxford, 2001).
\bibitem{Xing2006} J. Xing and K. S. Kim, Phys. Rev. E 74, 061911 (2006).
\bibitem{Stein1985} D. L. Stein, Proc. Natl. Acad. Sci. U.S.A. 82, 3670 (1985)
\bibitem{Zwanzig1988} R. Zwanzig, Proc. Natl. Acad. Sci. U.S.A. 85, 2029 (1988).
\bibitem{Nevo2005} R. Nevo, V. Brumfeld, R. Kapon, et al., EMBO Rep. 6, 482 (2005); D. Thirumalai and S. A. Woodson, Acc. Chem. Res. 29, 433 (1996).
\bibitem{Kim2002} S. H. Kim, W. R. Wang, and K. K. Kim, Proc. Natl. Acad. Sci. U.S.A. 99, 11611 (2002).
\bibitem{Wang2003} H. Wang, C. Peskin, and Elston, T., J. Theo. Biol. 221, 491 (2003).
\bibitem{Gopich2006} I. V. Gopich and A. Szabo, J. Chem. Phys. 124, 154712 (2006).
\bibitem{George1972} T. F. George and W. H. Miller, J. Chem. Phys. 57, 2458 (1972).
\bibitem{Kramers1940} H. Kramers, Physica 7, 284 (1940); P. Hanggi, P. Talkner, M. Borkovec, Rev. Mod. Phys. 62, 251 (1990).
\bibitem{Hamelberg2005} D. Hamelberg, T. Shen, and J. A. McCammon, J. Chem. Phys. 122, 241103 (2005); 125, 094905 (2006).
\bibitem{Popovych2006} N. Popovych, S. J. Sun, R. H. Ebright, et al., Nat. Struct. Mol. Biol. 13, 831 (2006).
\end{references}
\end{document}